\documentclass[twocolumn,floatfix,prl]{revtex4}%
\usepackage[dvipdfmx]{graphicx}%
\usepackage{amsmath}%
\usepackage{amsfonts}%
\usepackage{amssymb}
\usepackage{bm}
\usepackage[dvipdfmx]{color}
\usepackage[dvipdfmx]{xcolor}
\usepackage{here}
\usepackage{ulem}

\def\0{{ {\bm 0} }}

\allowdisplaybreaks[4]

\begin{document}

\title{Microscopic origin of period-four stripe charge-density-wave in kagome metal CsV$_3$Sb$_5$}
\author{Yuma Murata$^1$, Rina Tazai$^2$, Youichi Yamakawa$^1$, Seiichiro Onari$^1$, and Hiroshi Kontani$^1$}
\date{April 2025}

\begin{abstract}
The interplay between unconventional density waves and exotic superconductivity has attracted growing interest.
Kagome superconductors $A\rm{V}_3\rm{Sb}_5$ ($A = \rm{K}, \rm{Rb}, \rm{Cs}$) offer a platform for studying quantum phase transitions and the resulting symmetry breaking.
Among these quantum phases, the $4a_0$ stripe charge-density-wave (CDW) has been widely observed for $A=\rm{Rb}$ and $\rm{Cs}$ by scanning tunneling microscopy (STM) and nuclear magnetic resonance (NMR) measurements.
However, the microscopic origin of the $4a_0$ stripe CDW remains elusive, and no theoretical studies addressing this phenomenon have been reported so far.
In this paper, we propose a microscopic mechanism for the emergence of the $4a_0$ stripe CDW.  
We analyze the CDW instability in the 12-site kagome lattice Hubbard model with the $2\times2$ bond order driven by the paramagnon-interference mechanism by focusing on the short-range magnetic fluctuations due to the geometrical frustration of kagome lattice.
We reveal that the nesting vector of the reconstructed Fermi surface, formed by the $2\times 2$ bond order, gives rise to a $4a_0$-period CDW.
Remarkably, the obtained stripe CDW is composed of both the off-site hopping integral modulations and on-site potentials.
The real-space structure of the stripe CDW obtained here is in good qualitative agreement with the experimentally observed stripe pattern. 

\end{abstract}

\affiliation{
$^1$Department of Physics, Nagoya University,
Furo-cho, Nagoya 464-8602, Japan.
$^2$Yukawa Institute for Theoretical Physics, Kyoto University, Kyoto 606-8502, Japan
}
\sloppy

\maketitle

\section{I. INTRODUCTION}
The kagome lattice superconductors $A{\rm V_3Sb_5}~(A=\rm{K},~\rm{Rb},~\rm{Cs})$ exhibit a geometric frustration and a characteristic band topology, both of which induce various density wave instabilities, leading to a cascade of quantum phase transitions
\cite{Ortiz1, Trans_Xs}.
In all members of $A\rm{V_3Sb_5}$, a time-reversal symmetric $2\times2$ bond order (BO), which is the staggered modulations of the hopping integrals, has been observed at ambient pressure at $T\approx 78$--$102~\rm{K}$
\cite{Ortiz2, Ortiz3, Tstar_Rb} 
by scanning tunneling microscopy (STM) 
\cite{STM_Cs1, STM_Cs2, STM_Cs3, STM_Rb1, STM_Rb2, STM_K1, STM_K2}. 
This $2\times2$ BO exhibits two distinct patterns, referred to as Tri-Hexagonal and Star-of-David patterns
\cite{ARPES1}. 
They are presumably triple-$\bm{q}$ BO, 
which is expressed as the real number spatial modulations in the hopping integrals, $\delta t^b_{ij} = \delta t^b_{ji} = \rm{real}$
\cite{BO1_FRG, BO2_FRG, BO3_RPA, BO4_GL, BO5_PRG, BO6_GL, BO7_tazai}.
The emergence of the $2\times2$ BO is attributed to the nesting between different van Hove singularity (vHS) points. 
Below the BO transition temperature $T_{\rm{BO}}$, nodeless superconductivity emerges in the case of $A=\rm{Cs}$
\cite{SC1, SC2}, 
which is naturally explained by the BO fluctuation mechanism proposed in Ref.~\cite{BO7_tazai}.
Time-reversal symmetry breaking (TRSB) ($A=\rm{K},~\rm{Rb},~\rm{Cs}$) is also detected by STM
\cite{STM_K1, STM_Rb1}, 
$\mu$SR
\cite{MuSR1, MuSR2, MuSR3, MuSR4}, 
and field-tuned chiral transport study
\cite{Moll1}. 
TRSB in $A\rm{V_3Sb_5}$ is considered a signature of loop current (LC) order, 
which is inherently understood as the pure-imaginary modulations in the hopping integrals, $\delta t^c_{ij} = -\delta t^c_{ji} = \rm{imaginary}$
\cite{LC1_MFA, LC2_MFA, LC3_MFA, LC_FRG, LC4_tazai}.

In addition, the emergence of the $4a_0$ stripe charge-density-wave (CDW) has been attracting great attention.
$a_0$ denotes the distance between the next-nearest-neighbor V-sites in Fig.~\ref{fig:model}(a). 
The $4a_0$ stripe CDW has been observed for $A=\rm{Rb}, \rm{Cs}$ by STM
\cite{STM_Cs1, STM_Cs2, STM_Cs3, STM_Rb1, STM_Rb2}
and NMR
\cite{NMR}.
The transition temperature at ambient pressure is $T_{\rm stripe} \approx 35~\rm{K}$
\cite{Moll1, Moll2}.
This $4a_0$ stripe CDW is a multiple quantum phase transition that occurs within the $2\times2$ BO phase, where the $4a_0$ stripe CDW coexists with the $2\times2$ BO.
However, to our knowledge, no theoretical mechanism for the emergence of this stripe CDW has been proposed.

The $4a_0$ stripe CDW breaks inversion symmetry (IS) by coexisting with loop current order
\cite{eMChA_tazai}.  
The IS breaking is prominently observed as nonreciprocal phenomena in kagome metal CsV$_3$Sb$_5$.  
In the normal state, electronic magnetochiral anisotropy (eMChA) appears around $T \approx 35~\rm{K}$
\cite{Moll1}.  
A key feature is a large eMChA coefficient $\gamma_{\rm{eM}}$ in $\rho_{zz} = \rho^0_{zz}(1 + \gamma_{\rm{eM}} B_x j_z)$ and the anisotropy to be switched by a small magnetic field $B_z$.  In the superconducting state, the superconducting diode effect, where the critical current $I_c$ becomes nonreciprocal with respect to current direction, has been observed
\cite{SDE}.  
Near the quantum critical point of the $4a_0$ stripe CDW, the superconducting transition temperature $T_c$ is expected to increase.
These findings indicate that the $4a_0$ stripe CDW plays a central role in kagome superconductors.

Extensive theoretical studies have explored the rich quantum phases in kagome metals.  
Within the framework of the mean-field approximation (MFA), large off-site bare interactions $V$ are required to stabilize BO and loop current order
\cite{BO4_GL, LC1_MFA, LC2_MFA, LC3_MFA}. 
Instead introducing large $V$, it is found that the effective off-site interaction due to beyond-MFA mechanisms can also cause various quantum phase transitions
\cite{BMFA1, BMFA2, BMFA3, BMFA4, BMFA5, BMFA6, BMFA7, BMFA8, BMFA9, BMFA10, BMFA11, BMFA12, BO7_tazai, LC4_tazai, LC_FRG}. 
In kagome metals, it has been revealed that the mechanism of quantum interference of paramagnons,
which is an important beyond-MFA effect,
gives rise to the BO state
\cite{BO7_tazai}.
This quantum interference is significant in kagome metals because geometric frustration prohibits the freezing of paramagnons, leading to the SDW order.
In this mechanism, the BO originates from the effective intersite interactions derived solely from the on-site Hubbard interaction $U$.
Furthermore, it has been shown that the BO fluctuations can induce current order ($={\rm imaginary}~\delta t^c_{ij}$) that breaks time-reversal symmetry with odd parity
\cite{LC4_tazai}.  
However, the resulting CDWs exhibit $2a_0 \times 2a_0$ modulation, and no $4a_0$-periodic CDW has been derived.  

In this paper, we propose a microscopic mechanism for the emergence of the $4a_0$ stripe CDW 
at $T=T_{\rm{stripe}}\approx 35~\rm{K}$, inside the $2\times2$ BO phase below $T_{\rm{BO}}\approx 90~\rm{K}$.
To examine this mechanism, we construct a 12-site extended-kagome-lattice model with 
the static $2\times2$ BO.
By investigating CDW instability due to the paramagnon-interference mechanism, we reveal the development of a $4a_0$-period CDW. 
The nesting vector of  the emergent Fermi surface pocket around the $\Gamma$ point (Fig.~\ref{fig:model}(c)) corresponds to the wavevector of the $4a_0$ CDW.
Interestingly, in the obtained $4a_0$ stripe CDW, the largest components are the long-range hopping modulations across the hexagon on the kagome lattice, followed by the site potential modulations. 
This real-space structure of the stripe CDW is in good agreement with STM measurements, and it is expected to provide a microscopic basis for understanding nonreciprocal transports, such as the eMChA and superconducting diode effect
\cite{Moll1, SDE}.

\section{I\hspace{-1.2pt}I. RESULTS}
\subsection{A. 12-site extended-kagome-lattice model}

\begin{figure}[htb]
\includegraphics[width=0.99\linewidth]{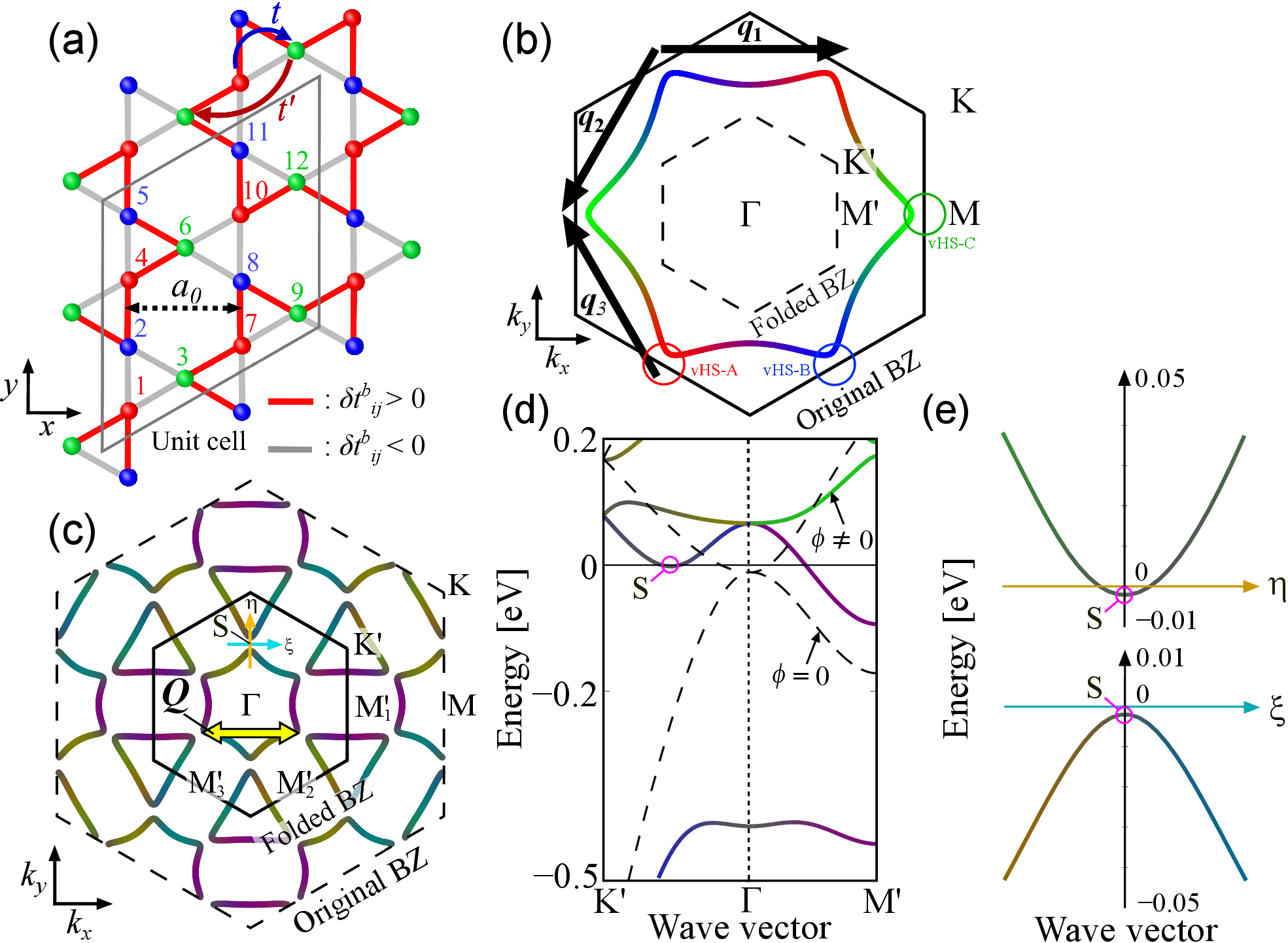}
\caption{
\textbf{Lattice structure, Fermi surface of 3-site and 12-site model, and band structure.}
(a) Kagome lattice structure of the vanadium layer and Star-of-David (SoD) BO pattern. 
The red and gray bonds mean the modulation at $+\phi$ and $-\phi$, respectively. 
The two arrows $t$ and $t'$ shown at the top of the unit cell represent the nearest- and third-nearest-neighbor hopping integrals, with strengths of $-0.5~\rm{eV}$ and $-0.08~\rm{eV}$, respectively.
(b) Fermi surface of 3 sites mode at $n=2.8$.
$\bm{q}_1,\bm{q}_2$ and $\bm{q}_3$ are the nesting vectors of the $3Q$ BO, connecting van-Hove singularity (vHS) points.
(c) Fermi surface of 12-site model under BO at $n=11.2$ (corresponding to $n=2.8$ in the 3-site model), $\phi=0.08~\rm{eV}$. 
$\bm{Q}$ is the new nesting vector by folded fermi surface.
(d) The colored solid lines show the band structure with SoD BO shown in (a), while the dashed lines correspond to the case with $\bm{\phi} =0$, both calculated at filling $n = 11.2$.
The band structure near the $\Gamma$ point is split into a 2-to-1 configuration due to the SoD BO.
(e) Band structures along the $\eta$ and $\xi$ paths with a saddle point at $S$ point in (c).   
By $3Q$ BO symmetry, saddle points also exist at the other corners of the hexagonal Fermi surface.  
}
\label{fig:model}
\end{figure}

Here, we introduce a 12-site extended-kagome-lattice model shown in Fig.~\ref{fig:model}(a).
The $4a_0$ stripe CDW emerges within the $2\times2$ BO phase, and the formation of the superlattice due to the $2\times2$ BO is considered to play a crucial role in this stripe CDW.
In our proposed mechanism for the emergence of the $4a_0$ stripe CDW, an analysis based on a 12-site model that incorporates the $2\times2$ BO is required.

This model consists of 12 $b_{3g}~(d_{XZ})$ orbitals, incorporating modulations of the hopping integral induced by the BO.
The kinetic term and BO term is given as
\begin{equation}
\begin{split}
    \hat{H}_0 &=\sum_{i,j,\sigma}\sum_{l,m}\Bigl(t_{i,j,l,m}+\delta t^b_{i,j,l,m}\Bigr) c_{i,l,\sigma}^{\dagger}c_{j,m,\sigma}\\
    &=\sum_{\bm{k},\sigma}\sum_{l,m} H^0_{lm}(\bm{k})c_{\bm{k},l,\sigma}^{\dagger}c_{\bm{k},m,\sigma}
\end{split}
\label{eq:H}
\end{equation}
where $l,m$ are the indices of the sublattice ($l,m=1,2,\cdots,12$), and $i,j$ are the indices of the unit cell positions. 
The nearest- and third-nearest-neighbor hopping integrals are $t=-0.5$ $\rm{eV}$ and $t'=-0.08$ $\rm{eV}$, respectively, and they reproduce well the Fermi surface obtained from first-principles calculations.
In the second line of Eq.~(1), we perform the Fourier transformation of the creation and annihilation operators as $c_{i,l,\sigma} = \frac{1}{\sqrt{N}} \sum_{\bm{k}} e^{i \bm{R}_i \cdot \bm{k}} c_{\bm{k},l,\sigma}$, where $\bm{R}_i$ denotes the position of the unit cell $i$.

The BO term $\delta t^b_{i,j,l,m}$ is the modulation of the hopping integral between $(i, l)$ and $(j, m)$ atoms by $2\times2$ BO.
Here, we incorporate a Star-of-David (SoD) type BO shown in Fig.~\ref{fig:model}(a), where the modulation amplitude ${\rm max}|\delta t_{i,j,l,m}|=\phi$ is set as $0.08~\rm{eV}$.
Theoretically, $\delta t^b_{i,j,l,m}$ represents the symmetry-breaking component of the self-energy and such a symmetry breaking is derived from the density-wave (DW) equation discussed later
\cite{DW}. 
From previous theoretical studies 
\cite{BO3_RPA, BO4_GL, BO5_PRG, BO7_tazai, LC4_tazai}, 
the wave vector of the BO $\bm{q}_n ~(n=1,~2,~3)$ corresponds to the nesting vectors connecting van-Hove singularities (vHS) shown in Fig.~\ref{fig:model}(b).
Since $T_{\rm BO} \gg T_{\rm stripe}$, we can safely study the mechanism of the stripe CDW by introducing constant $\delta t_{i,j,l,m}^b$.
The BO patterns such as the SoD and the TrH are characterized by these three wave vectors : the $3Q$ BO.
Here, we represent the BO for the wave vector $\bm{q}_n ~(n=1,~2,~3)$ as $\bm{\phi}=(\phi_1,\phi_2,\phi_3)$, 
and the corresponding BO form factors as 
$\bm{g}_{l,m}=(g^{(1)}_{l,m},g^{(2)}_{l,m},g^{(3)}_{l,m})$.
We introduce the simplified even-parity BO $g^{(n)}_{l,m}=g^{(n)}_{m,l}=\pm 1$ between the nearest-neighbor sites.
As only the nearest-neighbor bonds are taken into account, 
the unit-cell indices $i$ and $j$ can be omitted.
Then, $\delta t^b_{l,m}$ can be expressed as
\begin{equation}
\delta t^b_{l,m} = \bm{\phi} \cdot \bm{g}_{l,m}.
\label{eq:BO}
\end{equation}
For example, the BO form factor at $\bm{q}=\bm{q}_1$ , $g^{(1)}_{l,m}$, is finite on the nearest neighbor bonds perpendicular to $\bm{q}_1$, such as $(l,m)=(1,2), (2,4), (4,5), (5,1)\cdots$.
Here, we introduce the SoD BO as $g^{(1)}_{l,m}=+1~[-1]$ for 
$(l,m)=(2,4),(5,1),(7,8),(10,11)$ $[(1,2),(4,5),(8,10),(11,7)]$, as shown in Fig.~\ref{fig:model}(a).  
The other components $g^{(2)}_{l,m}$ and $g^{(3)}_{l,m}$ are immediately obtained from Fig.~\ref{fig:model}(a).
Thus, the amplitudes of the BO are expressed as $\bm{\phi}=(0.08,0.08,0.08)~\rm{eV}$.

The Fermi surface of the 3-site model at filling $n=2.8$ per 3-site unit cell is shown in Fig.~\ref{fig:model}(b). 
At this filling, the vHS energy is $E_{\rm vHS}\approx -0.012~\rm{eV}$ measured from the chemical potential.
(Note that $E_{\rm vHS}=-0.05\sim -0.1~{\rm eV}$ in $A\rm{V_3 Sb_5}$ ($A=\rm{Cs,~Rb~,K}$).)
The color of the Fermi surface represents the orbital dependence corresponding to the sublattice colors in Fig.~\ref{fig:model}(a). 
Here, red means the sublattices $1, 4, 7, 10$.
Also, blue and green correspond to the sublattices $2, 5, 8, 11$ and $3, 6, 9, 12$, respectively.
The red, blue, and green sublattices correspond to A, B, and C sublattices in the original 3-site kagome lattice model, respectively.  
This is a pure-type Fermi surface, where the Fermi surface near the vHS points is composed of a single sublattice A, B, or C.
For the 12-site model, the original 3-site Brillouin zone is reduced to the folded Brillouin zone. 
In addition, the pseudogap opens under the BO as shown in the Fig.~\ref{fig:model}(d). 
This 2-to-1 splitting of the bands near the $\rm{\Gamma}$ point is characteristic of the SoD BO: 1-to-2 splitting appears in the reversed SoD (TrH) BO 
\cite{CDW-pattern}.
The resulting folded Fermi surface under BO at $n=11.2$ per 12-site unit cell and $\phi=0.08 \rm{eV}$ is exhibited in Fig.~\ref{fig:model}(c). 
As we see in Figs.~\ref{fig:model}(d) and (e), both the hexagonal and triangular Fermi surfaces are hole pockets. 
As shown in Fig.~\ref{fig:model}(e), six saddle points exist between the nearest hexagonal and triangular pockets. 
Therefore, the yellow arrow in Fig.~\ref{fig:model}(c) works as the nesting vector, which is  responsible for the $4a_0$-period CDW.

\subsection{B. Linearized density-wave equation}

Kagome metals exhibit nonlocal orders, such as BO, which cannot be captured within mean-field approximations for on-site coulomb interaction $\hat{H}_U = \sum_{i} Un_{i\uparrow}n_{i\downarrow}$.
Therefore, in this study, we perform an analysis using the linearized density-wave (DW) equation incorporating vertex corrections (VCs) beyond mean-field approximation.
The linearized DW equation is given by
\begin{equation}
\begin{split}
    \lambda_{q}f^L_{q}(k) &=-\frac{T}{N}\sum_{p,M_1,M_2}I^{L,M_1}_q(k,p)\\
    &\times\bigl\{G(k)G(k+q)\bigr\}^{M_1,M_2}f_q^{M_2}(p),
\end{split}
\label{eq:DW}
\end{equation}
where $k\equiv (\bm{k},\epsilon_n)$, $p\equiv (\bm{p},\epsilon_m)$ ($\epsilon_n$, $\epsilon_m$ are fermion Matsubara frequencies), $L\equiv (l,l')$ and $M_i \equiv (m_i , m_i ')$ represent the pair of sublattice indices.
The eigenvalue $\lambda _{q}$ characterizes the instability of the density-wave (DW) state, and a phase transition occurs when $\lambda _{q}$ reaches 1.
The eigenstate $f^L_{q}(k)$, called form factor, gives the nature of the order and the associated symmetry breaking.
Thus, the state of $f^L_{q}(k)$ corresponding to the largest $\lambda _{q}$ is realized in the system. 
This $f^L_{q}(k)$ is proportional to the electron-hole condensation $-\sum_{\sigma} \big(\langle c^{\dagger}_{\bm{k+q},l,\sigma}c_{\bm{k},l',\sigma}\rangle-\langle c^{\dagger}_{\bm{k+q},l,\sigma}c_{\bm{k},l',\sigma}\rangle_0\big)$
, which represents the symmetry-breaking component of the self-energy.

As defined in Ref.~\cite{DW}, 
$I^{L,M}_q(k,p)$ is given by the Ward identity $\delta ^2 \Phi _{\rm{LW}}/\delta G_{l',l}(k)\delta G_{m,m'}(p)$ at $\bm{q}=\bm{0}$ , where $\Phi _{\rm{LW}}$ is the Luttinger-Ward function
\cite{LW}.
To drive experimental CDW naturally, the choice of the appropriate 4-point vertex correction $I^{L,M}_q(k,p)$ is crucial.
In approximations based on the Luttinger-Ward functional $\Phi_{\rm{LW}}$, various physical quantities such as particle number are guaranteed to be conserved, in accordance with the Baym-Kadanoff conserving approximation
\cite{BK}. 
This framework has the advantage of eliminating unphysical solutions.
If the bare interaction is taken as $I^{L,M}_q$, which corresponds to RPA, the relation $\lambda_q > \alpha_s$ is not satisfied when the electron-electron interaction is the on-site Coulomb interaction $U$.
Therefore, higher-order correlations are indispensable to reproduce the BO theoretically.

\begin{figure}[htb]
\includegraphics[width=0.99\linewidth]{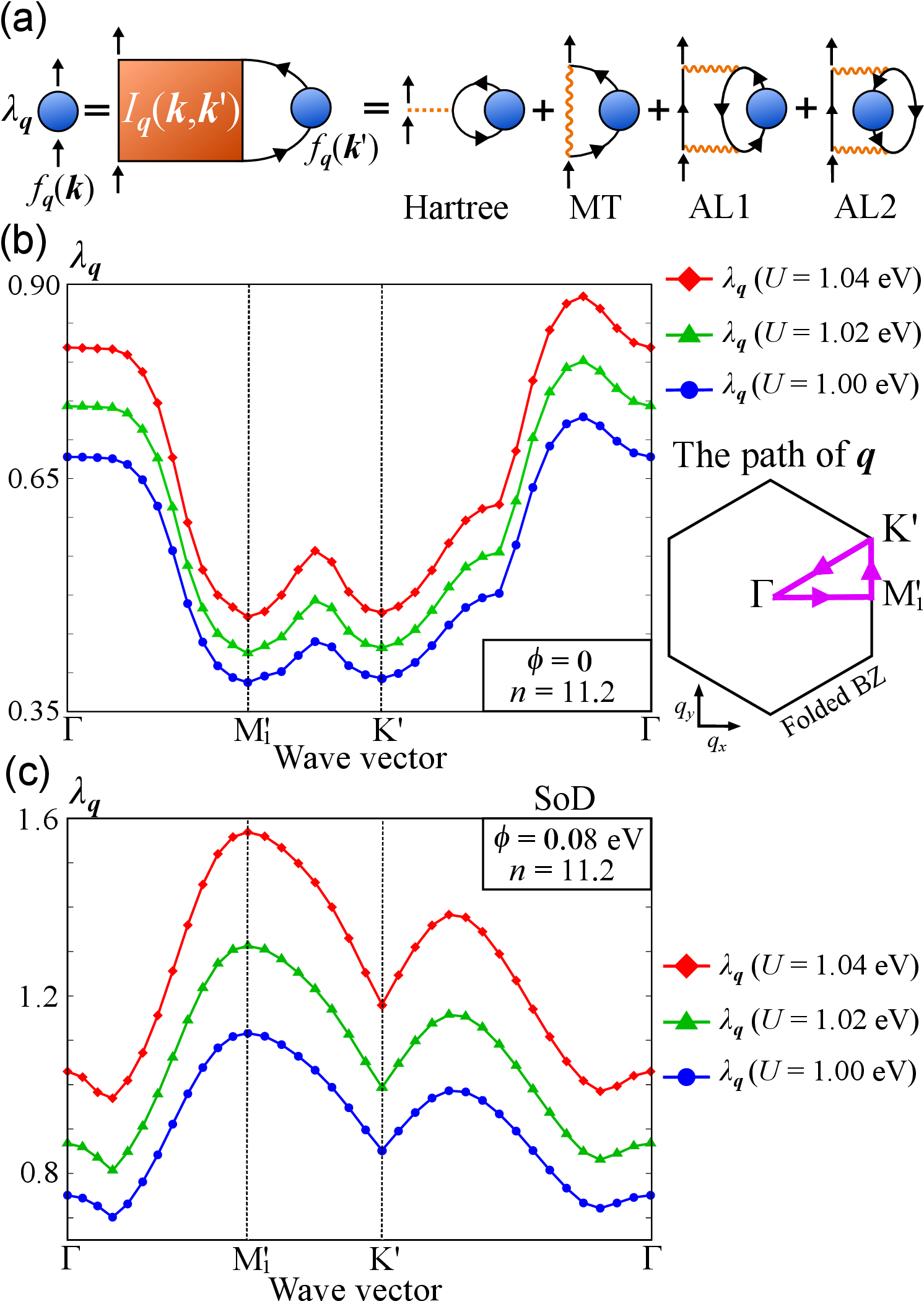}
\caption{
\textbf{Diagrammatic representation of linearized DW equation, the development of the eigenvalue corresponding to the $4a_0$ stripe CDW
}
(a) Diagrammatic representation of  Linearized DW equation. 
(b),(c) $\bm{q}$ dependence of the eigenvalue for $\phi=0$ and $\phi=0.08~\rm{eV}$, and the $\bm{q}$-space path.
Both analyses are based on the 12-site kagome lattice model.
The red, green, and blue lines represent the results for $U=1.00~\rm{eV}$, $U=1.02~\rm{eV}$, $U=1.04~\rm{eV}$ at $n = 11.2$. 
The $\bm{q}$-space path follows the folded Brillouin zone along the route $\rm{\Gamma} \rightarrow M'_1 \rightarrow K' \rightarrow \Gamma$.
All calculations are performed at $T = 0.01~\rm{eV}$.  
}
\label{fig:DW}
\end{figure}

In this study, we apply the one-loop approximation for $\Phi _{\rm{LW}}$.
As a consequence, we perform an analysis that incorporates not only the nonlocal Hartree term but also the vertex corrections given by the Maki-Thompson (MT) term and the Aslamazov-Larkin terms (AL1 and AL2) shown in Fig.~\ref{fig:DW}(a).
Two AL terms represent the interference mechanisms between two spin propagators with different momentum.

\subsection{C. Numerical results of $4a_0$ stripe CDW in kagome metals}

Here, to understand the $4a_0$ stripe CDW that emerges at $T_{\rm{stripe}} \approx 35~\rm{K}$,  
we analyze the 12-site model with the SoD BO introduced as the constant $\phi$ shown in Fig.~\ref{fig:model}(a),  
using the DW equation in Eq.~(\ref{eq:DW}).
Here, we treat the $2\times2$ BO as a constant order parameter.
This analysis is justified by the relationship $T_{\rm BO} \gg T_{\rm stripe}$.
In this case,  
the relation $\phi \gg I_{\rm{stripe}}$ 
($I_{\rm{stripe}}$ denotes the strength of the stripe CDW) is naturally expected.
By solving the self-consistent DW equation in Ref.~\cite{DW},  
a unified understanding of the multiple quantum transitions between the BO and stripe CDW can be obtained.

Figure~\ref{fig:DW}(b) shows the $\bm{q}$ dependence of the eigenvalue $\lambda _{\bm{q}}$ calculated for the 12-site model without BO ($\phi = 0$) at $n=11.2$ and $T=0.01~\rm{eV}$.  
The blue, green, and red lines represent $U=1.00~\rm{eV}$ ($\alpha_s=0.8$), $U=1.02~\rm{eV}$ ($\alpha_s=0.82$), and $U=1.04~\rm{eV}$ ($\alpha_s=0.84$), respectively 
(Note that $\alpha _s =0.8U$ in the present model.).
At $\bm{q}=\bm{0}$, the eigenvalue is threefold degenerate.
These three eigenvalues originate from the three $\rm{M}$ points in the original Brillouin zone, 
which correspond to $2a_0$-periodic modulations.
They are folded onto the $\Gamma$ point in the 12-site model.  
The form factors of these eigenvalues coincide with the $3Q$ BO obtained in the previous study based on the two-orbital model~\cite{BO7_tazai}.  
This $3Q$ BO is also reproduced by the first-principles-based 3D model for kagome metals $A\rm{V_3Sb_5}$~\cite{FP_onari}.  
In this analysis, the $2\times2$ BO are robustly obtained for all $A = \rm{K, Rb, Cs}$.
Although the peak deviates from $\bm{q}=\bm{0}$ in Fig.~\ref{fig:DW}(b), 
the $\bm{q}=\bm{0}$ order is expected to be favorable based on a realistic $3\rm{D}$ kagome model : see Ref.~\cite{FP_onari}.

Figure~\ref{fig:DW}(c) shows the $\bm{q}$ dependence of the eigenvalue under the SoD BO with $\phi = 0.08~\rm{eV}$.  
The blue, green, and red lines correspond to $U=1.00~\rm{eV}$ ($\alpha_s=0.9$), $U=1.02~\rm{eV}$ ($\alpha_s=0.92$), $U=1.04~\rm{eV}$ ($\alpha_s=0.94$), respectively ($\alpha _s =0.9U$).
For all $\alpha_s$, a robust peak appears at the $\rm{M'_1}$ point,  
which corresponds to a $4a_0$-periodic modulation.  
The present study naturally explains the $4a_0$ stripe CDW, 
which has been unsolved for years.
This $4a_0$ ordering vector corresponds to a new nesting vector of the Fermi surface shown in Fig.~\ref{fig:model}(c).  
The eigenvalue at the $\rm{M'}$ point is enhanced by this new nesting vector. 
By comparing the results for $\phi = 0$ and $\phi \neq 0$, It is noteworthy that the eigenvalue at the $\rm{M'}$ point increases by about $0.7–1.0$ for a fixed $U$ by introducing the SoD BO.
Furthermore, $\alpha_s$ increases by about $0.1$ due to the BO,  
indicating that the BO enhances spin fluctuations.  
These results indicate that the $2\times2$ BO below $T_{\rm BO}\approx 90~\rm{K}$ has a significant impact  
on the emergence of the $4a_0$ stripe CDW.  
This conclusion is also supported by the behavior of the irreducible susceptibility  
shown in Fig.~\ref{fig:xi} in Discussion section.  
Thus, we have successfully reproduced the multiple quantum phase transitions of the CDW phase in kagome metals.

Due to the rotational symmetry of the $3Q$ BO,  
$\lambda_{\bm{q}}$ takes the same value at $\rm{M'_1}$, $\rm{M'_2}$, and $\rm{M'_3}$ points.  
Therefore, the $4a_0$-periodic CDW can, in principle, emerge along three equivalent directions, although experimentally it appears only along one direction. 
This fact would be explained by the nematicity in the DW state,
due to the $\pi$-shift BO stacking 
\cite{BO5_PRG}
or the nematic BO+LC coexistence
\cite{LC4_tazai}.

\subsection{D. The real space-structure of the $4a_0$ stripe CDW from the form factor}

\begin{figure}[htb]
\includegraphics[width=0.99\linewidth]{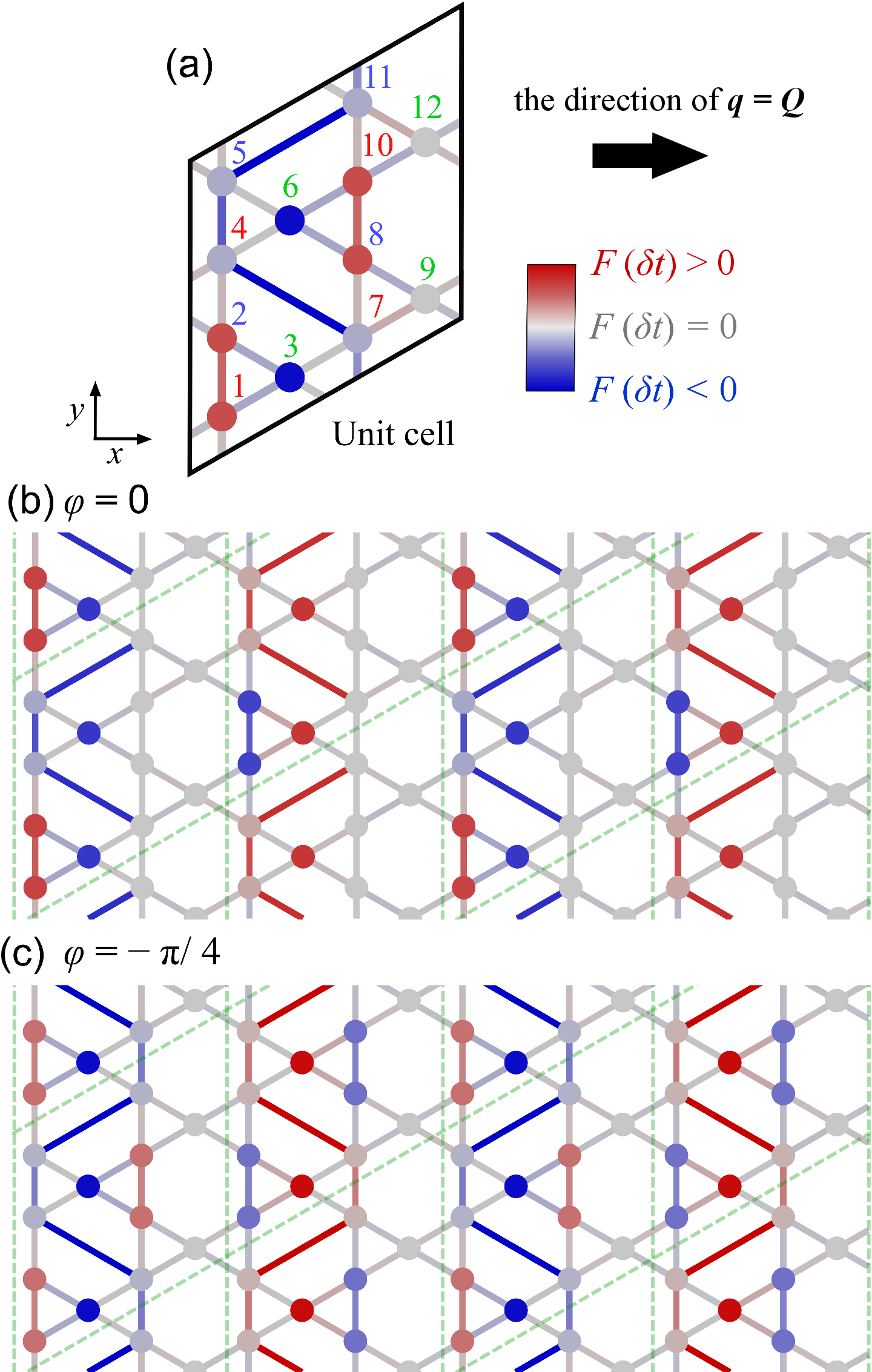}
\caption{
\textbf{Real-space representation of the form factor and the $4a_0$ stripe CDW structure}
(a) Real-space structure of the form factor. 
The colors of the bonds and sites represent the amplitude of modulation indicated by the color bar: blue is negative modulation, while red is positive modulation.
(b)-(c) show the structures with the constant phase at $\varphi =0,-\pi/4$, respectively.
The modulation amplitude is also indicated by the color bar shown in (a).
}
\label{fig:4a_0}
\end{figure}

We next focus on the real-space structure of the $4a_0$ stripe CDW.  
The eigenvalue at the $\rm{M'_1}$ point in Fig.~\ref{fig:DW}(c) corresponds to the CDW with a $4a_0$ periodicity, and we denote its ordering wave vector as $\bm{q}=\bm{Q}$.  
The vector $\bm{Q}$ corresponds to the nesting vector shown in Fig.~\ref{fig:model}(c).  
To obtain the real-space structure of this stripe CDW, we present the form factor in real-space representation at $\bm{q}=\bm{Q}$.  
The form factor is given by
\begin{equation}
    \tilde{F}^L_{\bm{q}}(\bm{R}_i)=\sum_{\bm{k}}F^L_{\bm{q}}(\bm{k})e^{i\bm{k}\cdot\bm{R}_i},
\label{eq:real}
\end{equation}
where $\bm{R}_i$ is a position of unit cell $i$.
$F^L_{\bm{q}}(\bm{k})\equiv f^L_{\bm{q}}(\bm{k}-\bm{q}/2)$ is the $\bm{q}/2$ shifted form factor, which ensures inversion symmetry around the origin in momentum space.
Figure~\ref{fig:4a_0}(a) shows the form factor $\tilde{F}^L_{\bm{Q}}(\bm{R})$,  
where the colors of sites and bonds represent the modulation strength.  
It is noteworthy that the strongest modulation appears in the components $L=(4,7)$ and $(5,11)$, which are the third-nearest-neighbor bonds across the kagome hexagon.  
The next strongest modulation is found in the site potentials of the $L=(3,3)$ and $(6,6)$ components. 
If we denote the modulation of the strongest components $|\tilde{F}^{4,7}_{\bm{Q}}|=|\tilde{F}^{5,11}_{\bm{Q}}|=F$,  
the site components take slightly smaller values
$|\tilde{F}^{3,3}_{\bm{Q}}|=|\tilde{F}^{6,6}_{\bm{Q}}|=0.95F$.
As discussed in Sec.III~B, the irreducible susceptibility corresponding to the components  
$L=(4,7)$ and $(5,11)$ is significantly enhanced at the $\rm{M'_1}$ point by the BO,  
which is in good agreement with the fact that $\tilde{F}^L_{\bm{Q}}$ exhibits the strongest modulation in these components.
Note that the quantitative results depend on model parameters.

From this real-space form factor,  
the real-space structure of the $4a_0$ stripe CDW can be obtained.  
The modulation $\delta t^{4a_0}_{i,j,L}$ is given by  
\begin{equation}
    \delta t^{4a_0}_{i,j,L}=2\tilde{F}^L(\bm{R}_i-\bm{R}_j)\cos\Bigl(\bm{q}\cdot\frac{\bm{r}_i+\bm{r}_j}{2}+\varphi\Bigl),
\label{eq:4a_0}
\end{equation}
where $\bm{r}_i = \bm{R}_i + \delta \bm{r}_i$ denotes the position of site $i$ within the unit cell at $\bm{R}_i$, $\delta \bm{r}_i$ represents the internal coordinate of the unit cell, and $\varphi$ is a constant phase.
Figs.~\ref{fig:4a_0}(b) and (c) display the modulations $\delta t_{i,j}^L$ with $\varphi=0,-\pi/4$, respectively.
For $\varphi=0$, the cosine maxima appear along the bonds such as $(1,2)$ and $(2,4)$,  
whereas for $\varphi=-\pi/4$, they appear along the $(7,8)$ and $(8,10)$ bonds.  
As discussed above, the long-range hopping integrals for $L=(4,7),~(5,11)$,  
as well as the site potentials for $L=(3,3),~(6,6)$,  
exhibit a clear $4a_0$-periodic modulation, which is qualitatively in good agreement
with the STM maps observed in experiments
\cite{STM_Cs1}.  
In practice, a background $2\times2$ BO is present, which in this case is the SoD-type BO.
The hopping integrals of the $L=(4,7)$ and $(5,11)$ components are located  
in positions connecting the Star-of-David clusters.  
The parameter $\varphi$ originates from the phase ambiguity in the linearized DW equation,  
and the selection of a particular $\varphi$ should be discussed  
in terms of coupling to the lattice and the stability of the higher-order Ginzburg--Landau free energy.
As shown in Figs.~\ref{fig:4a_0}(b) and (c),
global inversion symmetry is broken for $\varphi =0$,
while it is preserved for $\varphi =-\pi/4$.

\section{I\hspace{-1.2pt}I\hspace{-1.2pt}I. DISCUSSION}
\subsection{A. Effect of the $2\times2$ BO on the $4a_0$ Stripe CDW}
Our analysis demonstrates that a robust $4a_0$ CDW emerges under a model with a SoD-type BO of strong amplitude $\phi = 0.08~\rm{eV}$.  
Compared with the case of $\phi=0$, we find that the eigenvalue is significantly enhanced at the $\rm{M'_1}$ point, indicating that the BO is closely related to the emergence of the $4a_0$ stripe CDW. 
As shown in a later subsection, 
the corresponding component of the irreducible susceptibility is also enhanced at the $\rm{M'_1}$ point by the SoD BO.
The folded Fermi surface with the BO shown in Fig.~\ref{fig:model}(c) hosts a nesting vector corresponding to the $4a_0$ periodicity.  
This nesting vector connects the vicinity of the saddle points, where a large density of states accumulates.  
From the form factor corresponding to the $4a_0$ eigenvalue, we obtain the real-space structure of the $4a_0$ stripe CDW.  
To our knowledge, this is the first work to reveal its detailed real-space structure.  
A remarkable feature is that the hopping integrals across the kagome hexagon,  
namely the $L=(4,7)$ and $(5,11)$ components, exhibit the strongest modulation,  
showing a pronounced $4a_0$-periodic variation in these long-range bonds.
In these sublattice components, the $\rm{M'}$ point of the irreducible susceptibility is also enhanced by the SoD BO shown in Fig.~\ref{fig:xi}.
The site potentials in the $L=(3,3)$ and $(6,6)$ components also show strong modulation.  
The real-space structure of the $4a_0$ stripe CDW obtained in this study is in good qualitative agreement with the STM maps observed in experiments  
\cite{STM_Cs1}.  

\subsection{B. Effect of the $2\times2$ BO on the irreducible susceptibility}

\begin{figure}[htb]
\includegraphics[width=0.99\linewidth]{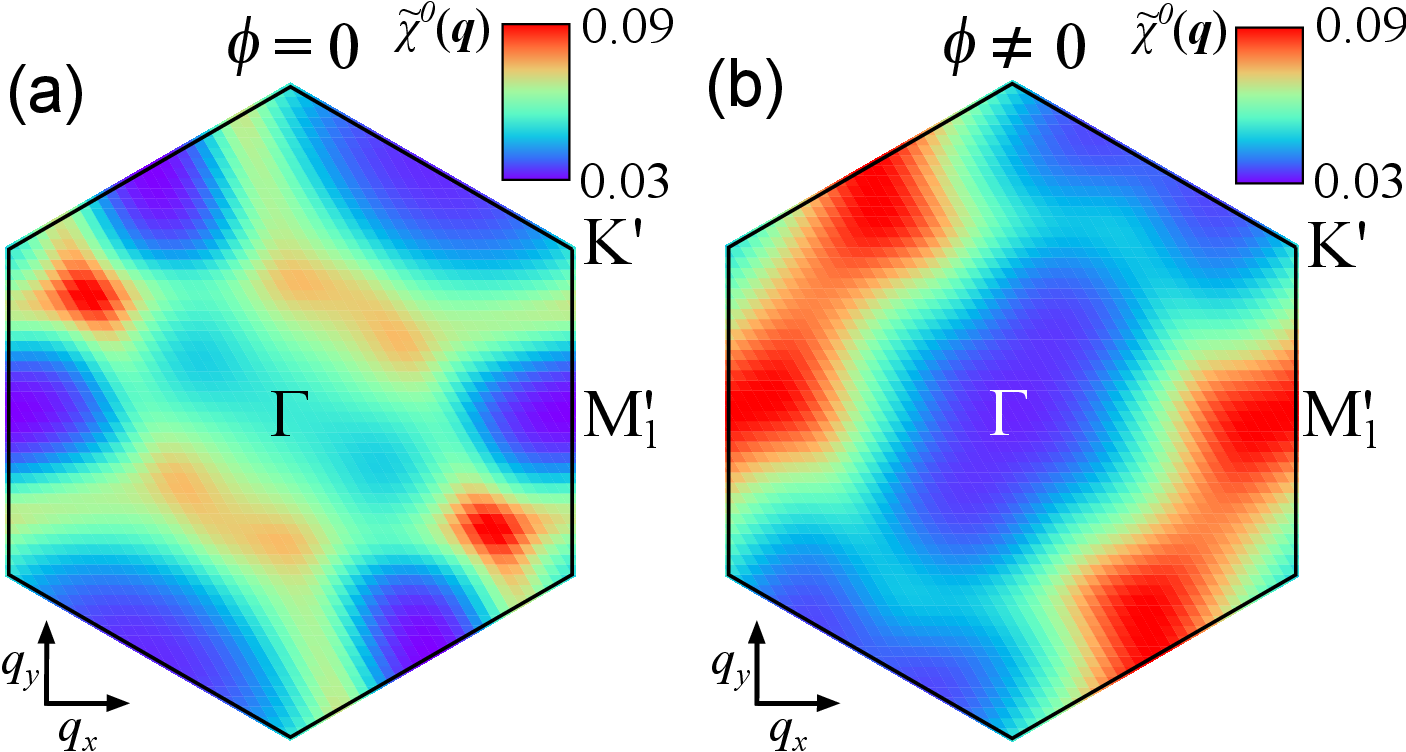}
\caption{
\textbf{Evolution of the irreducible susceptibility with BO}
(a) Irreducible susceptibility $\tilde{\chi} ^0 _{47,47}(\bm{q})$ of 12 sites model without BO ($n=11.2$). 
(b) $\tilde{\chi} ^0 _{47,47}(\bm{q})$ of 12 sites model with BO $\phi=0.08 ~\rm{eV}$ ($n=11.2$). The value at the $\rm{M'_1}$ point is obviously enhanced by incorporating BO. 
The expression of $\tilde{\chi}^0$ is given by Eq.~(\ref{eq:txi}).
}
\label{fig:xi}
\end{figure}

We demonstrate how the SoD BO modifies the irreducible susceptibility.
On the right-hand side of the DW equation in Eq.~(\ref{eq:DW}), 
there exists a term that behaves like an irreducible susceptibility,  
composed of the two Green’s functions between the four-point vertex and the form factor. 

The irreducible susceptibility is given by
\begin{equation}
    \chi^0_{ll',mm'}(q) = -\frac{T}{N} \sum_{k} G_{lm}(k+q)~ G_{m'l'}(k),
\label{eq:xi}
\end{equation}
where $q = (\bm{q}, \omega_n)$ and $k = (\bm{k}, \varepsilon_{m})$, $\omega_n = 2n\pi T$ is boson Matsubara frequency, and $\varepsilon_{m} = (2{m}+1)\pi T$ is fermion Matsubara frequency.
The non-interacting Green's function $G_{lm}(k)$ is given by the $(l,m)$ component of the matrix $\hat{G}(k) = \left( i \varepsilon_{m}\hat{1} - \hat{H}_0(\bm{k}) \right)^{-1}$, when $\hat{H}_0(\bm{k})$ is given in Eq.~(\ref{eq:H}) and $\hat{1}$ is the unit matrix.
To understand why the $4a_0$ stripe CDW obtained in the DW eq. (Fig.~\ref{fig:4a_0}), 
we introduce the following modified irreducible susceptibility, which corresponds to the electron-hole pair $\{G(k)G(k+q)\}$ in the DW equation.
\begin{equation}
\tilde{\chi}^0_{ll',mm'}(\bm{q})= -\frac{T}{N} \sum_{\varepsilon_m} {'}\sum_{\bm{k}} G_{lm}(k+q) ~G_{m'l'}(k),
\label{eq:txi}
\end{equation}
where we set $\sum '_{\varepsilon _m} = \sum_{\varepsilon_m}^{\pm \pi T}$ 
by considering the fact that both the MT and the AL vertex correction in $I_q(k,k')$, which appears in the DW eq., is large only for minimum Matsubara frequency $|\varepsilon_m|, |\varepsilon_{m'}| = \pi T$ 
at finite temperatures ($T \approx 100~\rm{K}$)
near the quantum criticality
\cite{BMFA12}. 
We present the irreducible susceptibility $\tilde{\chi}^0_{47,47}(\bm{q})$
calculated for the 12-site model without the BO ($\bm{\phi} = \bm{0}$) in Fig.~\ref{fig:xi}(a).
$(l,m)=(4,7)$ is third-nearest-neighbor pair, and the sublattices $4$, $7$  belongs to the sublattice A.
This sublattice pair $(4,7)$ represents the component for the largest modulation of the form factor shown in Fig.~\ref{fig:4a_0}(a).
A notable feature is that the introduction of the SoD BO significantly enhances the value of $\tilde{\chi}^0_{47,47}(\bm{q})$ at the $\rm{M'_1}$ point that exhibits a $4a_0$-period modulation : 
Fig.~\ref{fig:xi}(b) displays $\tilde{\chi}^0_{47,47}(\bm{q})$ with the SoD BO as $\phi=0.08~\rm{eV}$.

The long-range bond component $(5,11)$ of the form factor also shows the largest modulation as shown in Fig.~\ref{fig:4a_0}(a).
$(l,m)=(5,11)$ is also third-nearest-neighbor pair, and the sublattices $5$, $11$  belong to the sublattice B.
In accordance with this fact, $\tilde{\chi}^0_{5:11,5:11}$ is also enhanced at the $\rm{M'_1}$ point by the SoD BO.
The above irreducible susceptibilities are a representative ones that is strongly enhanced at the $\rm{M'_1}$ point by the SoD BO.  
These irreducible susceptibilities are considered to cause the strong enhancement of the eigenvalue at the $\rm{M'_1}$ point after the BO,  
and to generate the strongly modulated components of the form factor.

\subsection{C. $4a_0$ stripe CDW formation under TrH bond order}

\begin{figure}[htb]
\includegraphics[width=0.99\linewidth]{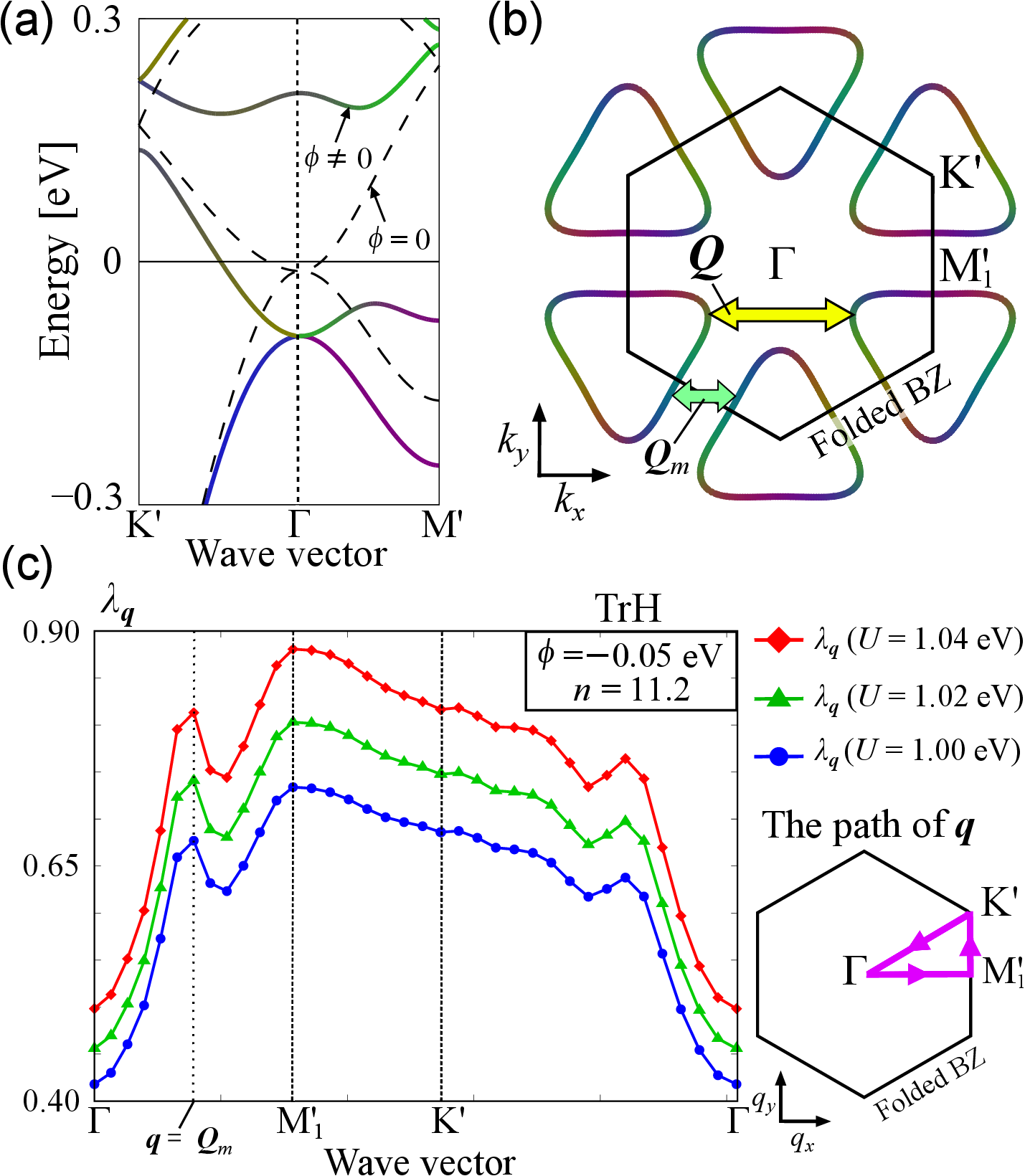}
\caption{
\textbf{Band structure, Fermi surface, and $\bm{q}$ dependence of the eigenvalues under the TrH BO.}
(a) Band dispersion under the TrH BO for $\phi = 0.05~\mathrm{eV}$ and $n = 11.2$.
(b) Corresponding Fermi surface. 
$\bm{Q}$ denotes the nesting vector corresponding to the $4a_0$ periodicity.
(c) $\bm{q}$ dependence of the eigenvalues under the TrH BO for $\phi = -0.05~\mathrm{eV}$ and $n = 11.2$.
The local maximum at $\bm{q} = \bm{Q}_m$ correspond to the nesting vectors $\bm{Q}_m$ indicated in (b).
}
\label{fig:TrH}
\end{figure}

The band splitting at the $\Gamma$ point differs between the SoD ($\phi > 0$) BO and TrH BO ($\phi < 0$).
In the SoD BO state,  the threefold degeneracy at the $\Gamma$ point at $E=E_{\rm vHS}$ splits into a doublet at $E_{\rm vHS}+2|\phi|-\Delta \mu$ and a singlet at $E_{\rm vHS}-4|\phi|-\Delta \mu$.
Here, $\Delta \mu$ is the shift of the chemical potential due to the $2\times2$ BO.
Conversely, in the TrH BO state, it splits into a singlet at $E_{\rm vHS}+4|\phi|-\Delta \mu$ and a doublet at $E_{\rm vHS}-2|\phi|-\Delta \mu$
\cite{CDW-pattern}.
In the present theory, the $4a_0$ stripe CDW is caused by the reconstructed Fermi surfaces near the split vHS points. 
Fig.~\ref{fig:TrH}(a) shows the band dispersion under the TrH BO at $\phi = -0.05~\rm{eV}$.

Here, we show the analysis of the CDW incorporating the TrH-type BO. 
For the TrH case, the eigenvalue of the DW equation exhibits a peak around the $4a_0$-periodic wave vector when $\phi \leq -0.04~\rm{eV}$, which is about half the magnitude of that in the SoD case.
Fig.~\ref{fig:TrH}(b) displays the Fermi surface for $\phi = -0.05~\rm{eV}$, where the nesting vector $\bm{Q}$  matches the $4a_0$-periodic ordering wave vector.
The $\bm{q}$ dependence of the eigenvalues under the TrH BO at $\phi = -0.05~\rm{eV}$, $n = 11.2$, $T = 0.01~\rm{eV}$ is shown in Fig.~\ref{fig:TrH}(c). 
The blue, green, and red lines correspond to $U = 1.00~\rm{eV}$ ($\alpha_s = 0.82$), $U = 1.02~\rm{eV}$ ($\alpha_s = 0.84$), and $U = 1.04~\rm{eV}$ ($\alpha_s = 0.85$), respectively.
As in the SoD case, a robust peaks emerge at the $\rm{M'_1}$ point for all $\alpha_s$, corresponding to the $4a_0$-periodic modulation.
The additional peak at $\bm{q} = \bm{Q}_m$ corresponds to the nesting vectors $\bm{Q}_m$ in Fig.~\ref{fig:TrH}(b).

Finally, we compare the experimental pseudo CDW gap with the theoretical BO gap in the present study. 
According to ARPES and optical conductivity measurements,
the pseudo gap $2\Delta _{\rm pg}$ inside the BO phase reaches $\sim 0.14~\rm{eV}$ and $\sim 0.17~\rm{eV}$, respectively
\cite{gap_ARPES, gap_opt}.
We obtained the clear peak at the $4a_0$-periodic wave vector by incorporating the SoD-type BO [Fig.~\ref{fig:DW}(c)], where $\phi = 0.08~\rm{eV}$.
A peak starts to develop around the $4a_0$-periodic wave vector at $\phi = 0.07~\rm{eV}$ in the SoD case, and the theoretical BO gap $2\Delta = 6|\phi|$ is $0.42~\rm{eV}$ at $\phi =0.07~\rm{eV}$. 
For the TrH BO parameter $\phi = -0.04~\rm{eV}$, a peak starts to develop at the $\mathrm{M'_1}$ point, and the corresponding theoretical BO gap is $2\Delta = 0.24~\mathrm{eV}$.
This gap is about half that in the SoD case. 
Although $2\Delta$ is still larger than the experimental $2\Delta_{\rm pg}$, its magnitude is of the same order.
It is expected to be improved by a more quantitative analysis based on a two-orbital model ($d_{XZ}+d_{YZ}$).

\section{I\hspace{-1.2pt}V. SUMMARY}
In this paper, we have proposed a microscopic mechanism for the emergence of the $4a_0$ stripe CDW that develops at $T_{\rm{stripe}} \approx 35~\rm{K}$ within the $2\times2$ BO phase below $T_{\rm{BO}} \approx 90~\rm{K}$.
Our analysis highlights that the $2\times2$ BO plays an essential role in the formation of the $4a_0$ stripe CDW.
We have shown that the Fermi-surface reconstruction due to the $2\times2$ BO induces the strong electron correlations and the emergence of the $4a_0$ stripe CDW.
The key for this phase transition is the paramagnon-interference mechanism.

To verify this mechanism, we constructed the 12-site extended kagome-lattice model incorporating the static $2\times2$ BO and analyzed it by the linearized DW equation.
As a result, we find that the $4a_0$-periodic CDW naturally emerges, which, to our knowledge, is demonstrated for the first time.
The realization of the $4a_0$ CDW originates from the new nesting vector formed on the folded Fermi surface shown in Fig.~\ref{fig:model}(c), and its eigenvalue is enhanced by the Aslamazov–Larkin (AL) term representing the paramagnon-interference mechanism.
Furthermore, the real-space structure of the $4a_0$ stripe CDW, obtained from the form factor of the DW equation, is in good qualitative agreement with the $4a_0$ stripe CDW observed by STM experiments~\cite{STM_Cs1}.

The present study paves the way for a detailed analysis of nonreciprocal phenomena based on the form factor obtained in this study,
like the electronic magneto-chiral anisotropy 
\cite{Moll1} 
and the superconducting diode effect
\cite{SDE}.
As a future problem, more quantitative discussion would be achieved by analyzing a two-orbital ($d_{XZ}+d_{YZ}$) kagome lattice model introduced in Ref.~\cite{LC5_tazai, QPI_nakazawa}, 
which is expected to clarify the material dependence of the stripe CDW.

\section{ACKNOWLEDGMENTS}
We are grateful to D. Inoue and S. Nakazawa for useful discussions.
This study has been supported by Grants-in-Aid for Scientific Research from MEXT of Japan (JP24K00568, JP24K06938, JP23K03299, JP22K14003) and a Grant-in-Aid for Transformative Research Areas (A) “Correlation Design Science” (KAKENHI Grant No. JP25H01246 and JP25H01248) from JSPS of Japan.

\appendix
\section{APPENDIX A: DERIVATION OF THE DW EQUATION}
\renewcommand{\theequation}{A.\arabic{equation}}
\setcounter{equation}{0}

\begin{figure}[htb]
\includegraphics[width=0.99\linewidth]{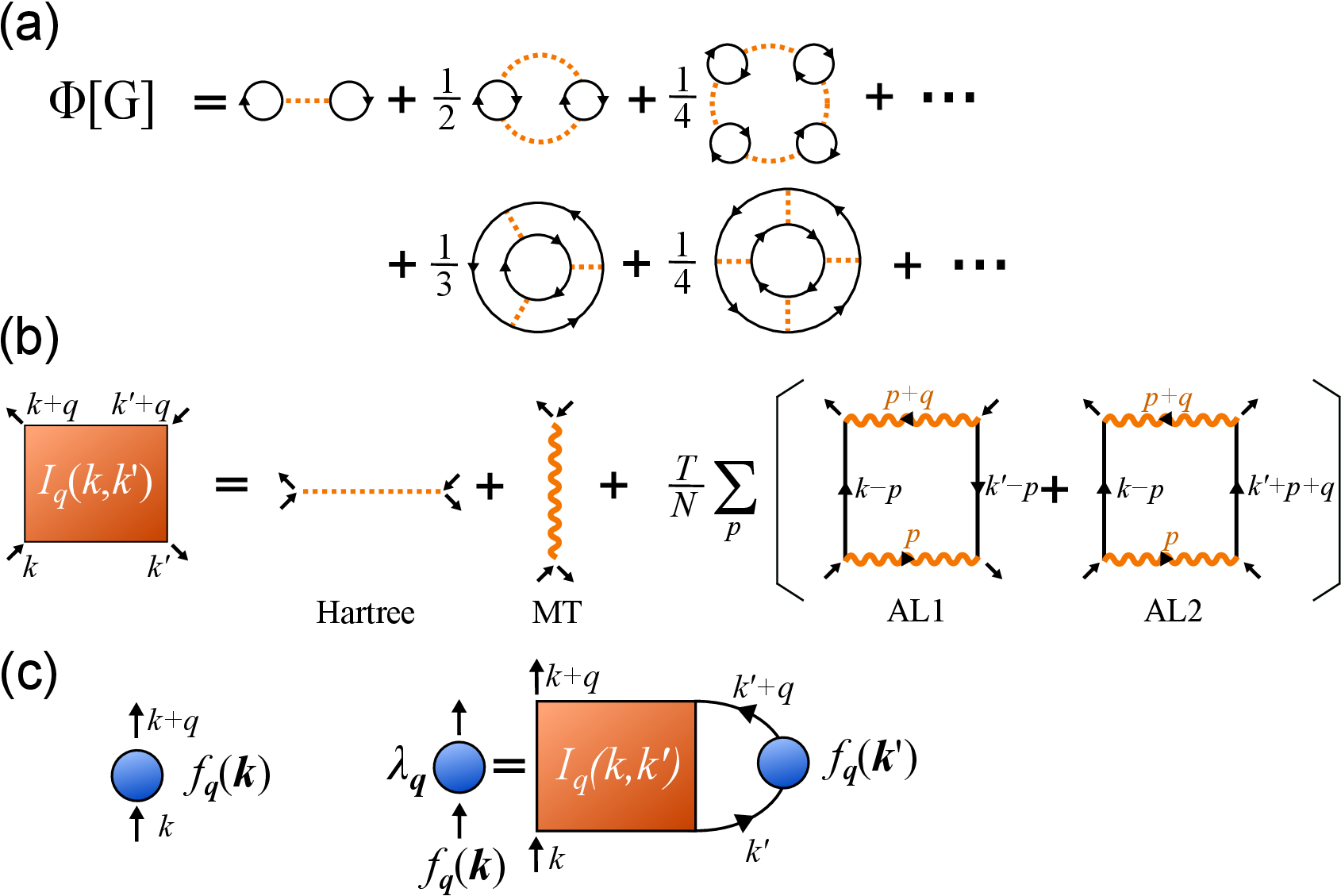}
\caption{
\textbf{Diagrammatic representations of the Luttinger–Ward potential $\Phi [G]$, 
the 4-point vertex $I$, and the DW equation.}
(a) Diagrammatic representation of the Luttinger–Ward potential $\Phi [G]$ 
under the one-loop approximation.
(b) The 4-point vertex $I$ obtained from $\Phi [G]$.
The first term represents the Hartree term, 
the second term corresponds to the Maki–Thompson (MT) term, 
and the third and fourth terms correspond to the Aslamazov–Larkin (AL) terms.
(c) Diagrammatic representation of the form factor $f_q(k)$ 
and the DW equation.
}
\label{fig:LW}
\end{figure}

In this section, we derive the expression of the 4-point vertex $I$ in order to obtain the DW equation. 
The 4-point vertex $I$ is defined as $\delta ^2 \Phi _{\rm{LW}}/\delta G_{l',l}(k)\delta G_{m,m'}(p)$, where $\Phi_{\rm{LW}}$ is the Luttinger–Ward functional~\cite{LW}. 
Here, we apply the one-loop approximation to $\Phi_{\rm{LW}}$~\cite{DW}. 
Under this approximation, $\Phi_{\mathrm{LW}}$ is represented by the ladder and bubble loop diagrams shown in Fig.~\ref{fig:LW}(a) for single-orbital Hubbard models.
Its analytic expression is 
\begin{equation}
\begin{split}
    \Phi [G] =&~T\sum_{q} {\rm Tr}\Bigg\{\frac{3}{2}{\rm ln}(\hat{1}-U^s\hat{\chi} _q^0)+\frac{1}{2}{\rm ln}(\hat{1}-U^c\hat{\chi} _q^0)\Bigg\} \\
    &+\frac{T}{4}\sum_q {\rm Tr}\big\{(U^s\hat{\chi} _q^0)^2+(U^c\hat{\chi} _q^0)^2 \big\} \\
    &+T\sum _q {\rm Tr} \Bigg\{\frac{3}{2}U^s\hat{\chi} _q^0+\frac{1}{2}U^c\hat{\chi} _q^0 \Bigg\},
\end{split}
\label{eq:LW}
\end{equation} 
where the trace ${\rm Tr}$ indicates the summation over sublattices, 
and the second term represents a $O(U^2)$ correction that is necessary to remove double counting of diagrams. 
Since only the on-site Coulomb interaction is considered, we set $U^s=-U^c=U$. 

By taking the second derivative of $\Phi_{\rm{LW}}$, 
the irreducible 4-point vertex $I$ is obtained as 
\begin{equation}
\begin{split}
    I &_{q}^{ll',mm'}(k,k') = -\frac{3}{2}V^s_{lm,l'm'}(k-k')-\frac{1}{2}V^c_{lm,l'm'}(k-k') \\
    &+\frac{T}{N}\sum_{b=s,c}\sum_{p}\sum_{n_1 n_2 n'_1 n'_2} \frac{a^b}{2}
    V^b_{m' n'_2,l'n_1}(p+\bm{q})V^b_{ln_2,m n'_1}(p) \\
    &~\times G_{n_1 n_2}(k-p)G_{n'_1 n'_2}(k'-p) \\
    &+\frac{T}{N}\sum_{b=s,c}\sum_{p}\sum_{n_1 n_2 n'_1 n'_2} \frac{a^b}{2}
    V^b_{n'_1 m,l'n_1}(p+\bm{q})V^b_{ln_2,n'_2 m'}(p) \\
    &~\times G_{n_1 n_2}(k-p)G_{n'_1 n'_2}(k'+p+\bm{q}) \\
    &+O(U^2),
\end{split}
\label{eq:I}
\end{equation} 
where $a^{s(c)}=3~(1)$ and $V^b$ denotes the interaction in the $b$ channel, 
given by $\hat{V}^b=U^b+U^b\hat{\chi} ^b_q U^b$. 
The fluctuation in the $b$ channel is expressed as 
$\hat{\chi} ^b_q=\hat{\chi} ^0_q ~(1-U^b \hat{\chi} ^0_q)^{-1}$. 

The first and second terms in Eq.~(\ref{eq:I}) correspond to the Maki–Thompson (MT) terms with first-order fluctuations, 
while the third and fourth terms correspond to the Aslamazov–Larkin (AL) terms with second-order fluctuations. 
The AL term represents the interference mechanism between quantum fluctuations.

By including the nonlocal contributions beyond the mean-field approximation through $I$, 
the DW equation is obtained as 
\begin{equation}
\begin{split}
    \lambda_{q}f^{ll'}_{q}(k) &=\frac{T}{N}\sum_{k',m_2,m'_2}K^{ll',m_2 m'_2}_q(k,k')f_q^{m_2 m'_2}(k'),
\end{split}
\label{eq:DWK}
\end{equation} 
where $K$ denotes the kernel of the DW equation, 
expressed as 
\begin{equation}
\begin{split}
    K^{ll',m_2 m'_2}_q(k,k')=&-\sum_{p,m_1,m'_1}I^{ll',m_1 m'_1}_q(k,k') \\
    &\times G_{m_1 m_2}(k'+\bm{q})G_{m'_2 m'_1}(k).
\end{split}
\label{eq:K}
\end{equation}
When the eigenvalue $\lambda_q$ of the linearized DW equation (Eq.~(\ref{eq:DWK})) reaches unity, 
it signals a phase transition. 
The corresponding form factor $f_q(k)$ represents the symmetry breaking of the self-energy. 
The form factor $f_q(k)$ obtained from the DW equation provides detailed information about the ordered state. 

Through the DW equation with MT and AL vertex corrections, 
the evolution of various effective off-site interactions can be explained without the need for introducing large off-site bare interaction. 
Due to this theoretical advantage,
the present theory has been naturally applied to novel quantum phase transitions
~\cite{BMFA3, BMFA6, BMFA7, BMFA8, BMFA11, BMFA12, BO7_tazai, LC4_tazai}. 

Higher-order diagrams beyond the mean-field approximation, such as MT and AL terms, are also included in the functional renormalization group (fRG) method. 
The fRG method has also explained nematic and smectic BOs~\cite{BMFA5, fRG}. 
This fact implies that diagrams beyond the MT and AL terms are not essential for describing BO formation.

\section{APPENDIX B: DW EQUATION ANALYSIS FOR $\phi=0.02\sim0.06~\rm{eV}$}

\begin{figure}[htb]
\includegraphics[width=0.99\linewidth]{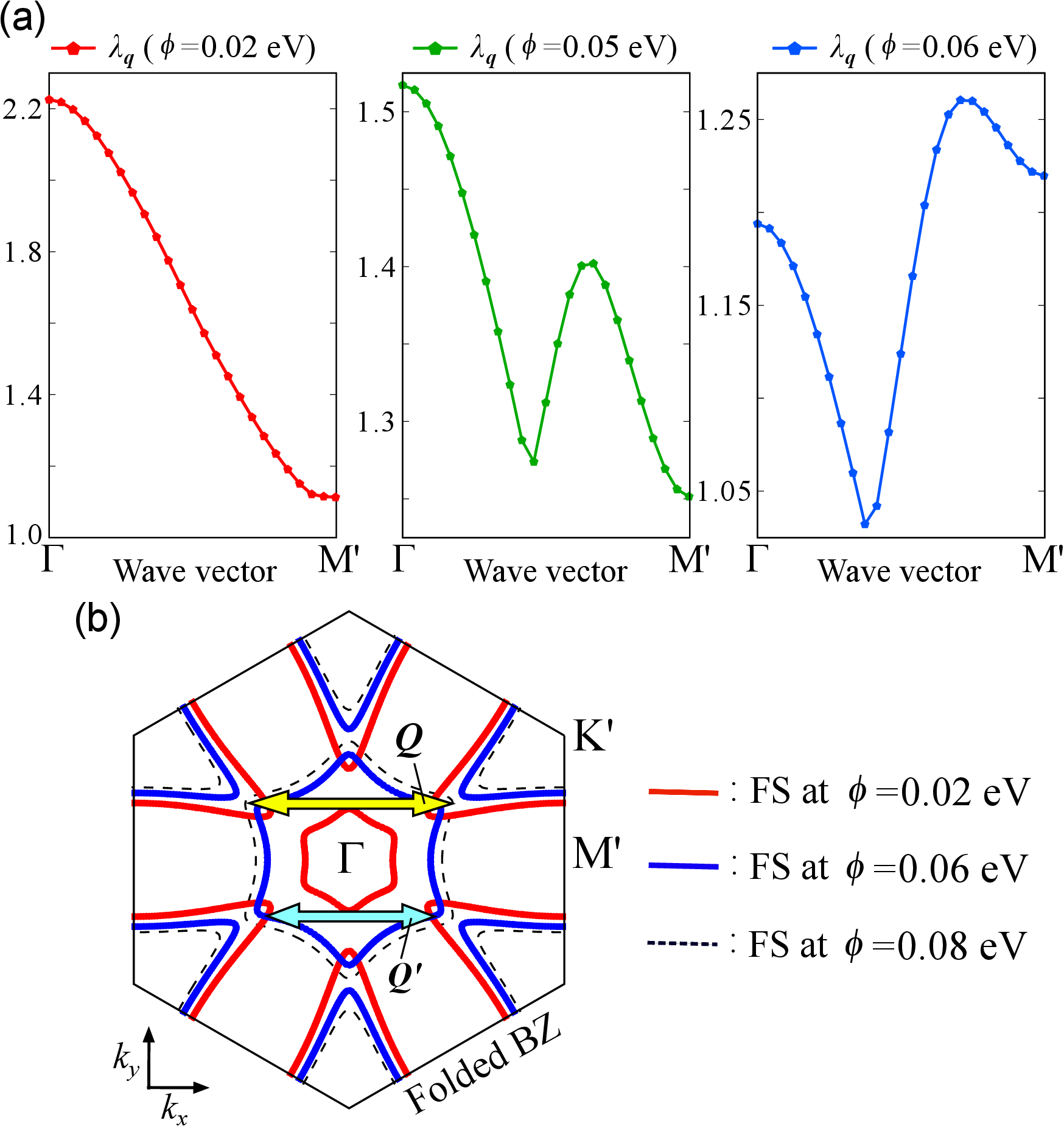}
\caption{
\textbf{$\bm{q}$ dependence of the eigenvalues for different values of $\phi$.
}
(a) $\bm{q}$ dependence of the eigenvalues for 
$\phi = 0.02~\rm{eV}$, $0.05~\rm{eV}$, and $0.06~\rm{eV}$. 
The $\bm{q}$ path is taken from the $\Gamma$ point to the $\rm{M'}$ point. 
All calculations were performed at $T = 0.01~\rm{eV}$ and $\alpha_s = 0.9$, 
with $U = 1.20$, $1.15$, and $1.10~\rm{eV}$, respectively. 
(b) Fermi surfaces for different values of $\phi$. 
The red and blue solid lines represent the Fermi surfaces 
for $\phi = 0.02~\rm{eV}$ and $\phi = 0.06~\rm{eV}$, respectively, 
while the dashed line corresponds to $\phi = 0.08~\rm{eV}$. 
The nesting vectors $\bm{Q'}$ and $\bm{Q}$ correspond to the eigenvalue peaks 
for $\phi = 0.06~\rm{eV}$ and $\phi = 0.08~\rm{eV}$, respectively.
}
\label{fig:BOdep}
\end{figure}

In this study, we solved the DW equation for the 12-site kagome-lattice model 
under the SoD BO with a large $\phi = 0.08~\rm{eV}$ 
and obtained a $4a_0$-periodic CDW. 
Here, we present the $\bm{q}$ dependence of the eigenvalues for other values of $\phi$. 
Figure~\ref{fig:BOdep}(a) shows the results along the $\Gamma$–$\mathrm{M'}$ path 
for $\phi = 0.02~\rm{eV}$, $0.05~\rm{eV}$, and $0.06~\rm{eV}$. 
All calculations were performed at $T = 0.01~\rm{eV}$ with $\alpha_s = 0.9$. 
For $\phi = 0.02~\rm{eV}$, no finite-$\bm{q}$ CDW develops. 
At $\phi = 0.05~\rm{eV}$, the eigenvalue at finite $\bm{q}$ starts to grow, 
and at $\phi = 0.06~\rm{eV}$, an incommensurate CDW emerges.
The magnitude of $\phi$ in real materials remains unclear.

Figure~\ref{fig:BOdep}(b) shows the Fermi surfaces for $\phi = 0.02~\rm{eV}$ 
and $\phi = 0.06~\rm{eV}$ (solid lines). 
The dashed Fermi surface corresponds to $\phi = 0.08~\rm{eV}$, 
where the eigenvalue peak appears at the $\mathrm{M'}$ point 
[Fig.~\ref{fig:DW}(c)]. 
The blue nesting vector $\bm{Q}'$ shown in Fig.~\ref{fig:BOdep}(b) 
corresponds to the eigenvalue peak for $\phi = 0.06~\mathrm{eV}$. 

These results indicate that the Fermi-surface geometry, 
which is strongly influenced by the BO, 
plays a crucial role in determining the CDW instability. 
Since the present model is a single-orbital ($d_{XZ}$-orbital) kagome model, 
more quantitative results are expected from a two-orbital 
($d_{XZ}+d_{XY}$) model that provides a more realistic Fermi surface.

\section{APPENDIX C: SPATIAL COMPONENTS OF THE FORM FACTOR IN THE $4a_0$ STRIPE CDW}

\begin{figure}[htb]
\includegraphics[width=0.99\linewidth]{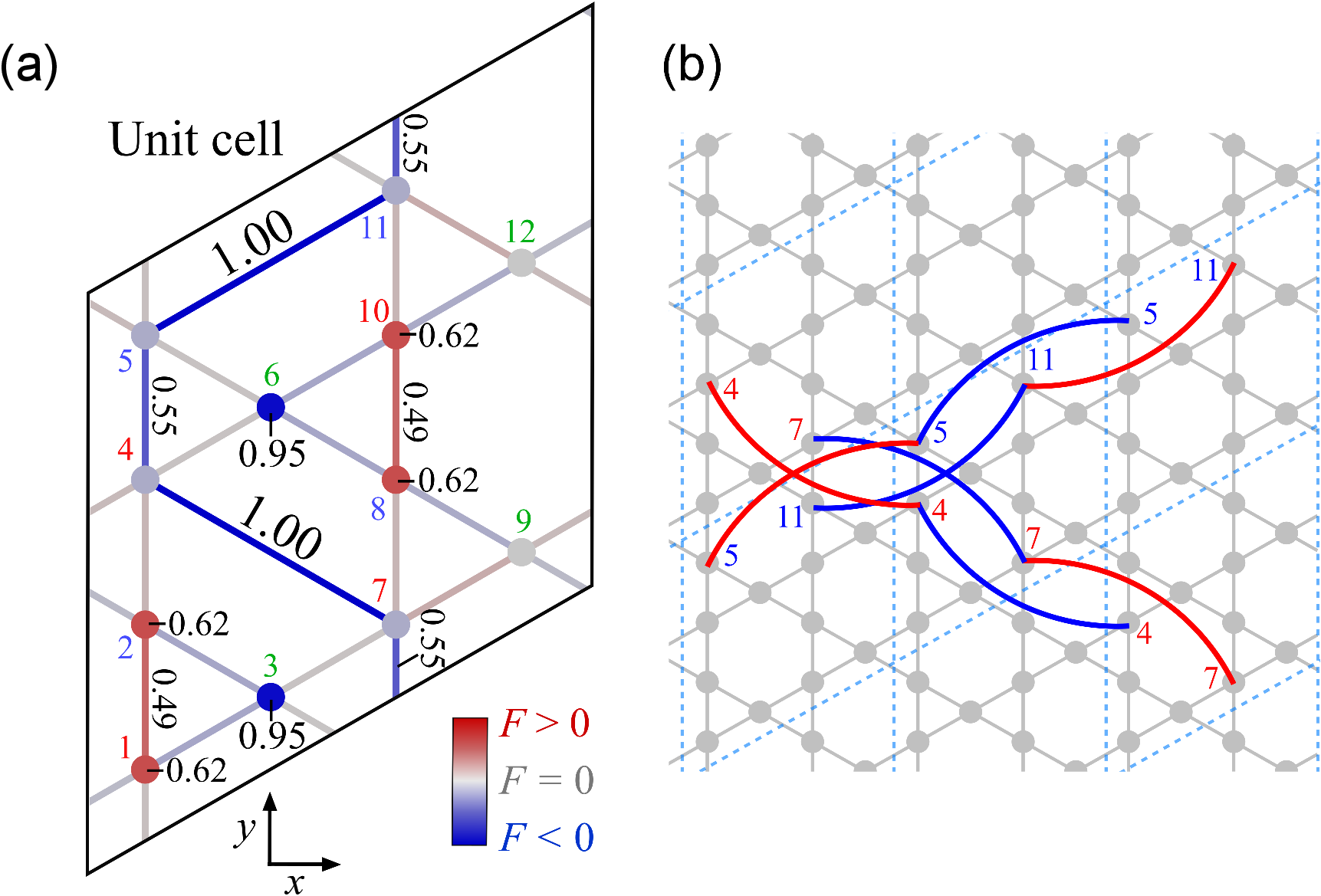}
\caption{
\textbf{Representative relative intensities of the real-space form factor 
$F_{\bm{Q}}^{L}(\bm{R})$.
}
(a) Relative intensities of various components normalized by the maximum modulation for $L = (4,7)$ and $(5,11)$, which are set to unity. 
Components with relative intensities smaller than $0.3$ are omitted. 
(b) Form factors $F_{\bm{Q}}^{L}(R)$ for the long-range bond components that extend over the unit cell, 
$L = (4,4)$, $(5,5)$, $(7,7)$, and $(11,11)$, whose relative intensities are $\pm 0.92$.
}
\label{fig:Re}
\end{figure}

From the form factor $f_{\bm{q}}(k)$, we obtained the real-space structure 
of the $4a_0$ stripe CDW. 
Here, we present representative values of the real-space form factor 
$F_{\bm{Q}}^{L}(\bm{R})$, where $\bm{Q}$ is the ordering vector corresponding to the $4a_0$ periodicity. 

The strongest modulations appear in the long-range bond components $L = (4,7)$ and $(5,11)$. 
The relative intensities shown in Fig.~\ref{fig:Re}(a) for the bonds and sites 
are normalized with respect to $F_{\bm{Q}}^{4,7} = F_{\bm{Q}}^{5,11} = 1$. 
The next strongest modulations are the site potentials 
$(3,3)$ and $(6,6)$, with $F_{\bm{Q}}^{3,3} = F_{\bm{Q}}^{6,6} = -0.95$. 
Other representative components are the site potentials $(1,1)$, $(2,2)$, $(8,8)$, and $(10,10)$ 
(with relative intensity $+0.62$), 
followed by the nearest-neighbor bond components $(4,5)$, $(7,11)$ (relative intensity $-0.55$), 
and $(1,2)$, $(8,10)$ (relative intensity $+0.49$). 
In addition, longer-range bond components that extend over the unit cell, such as $(4,4)$, $(5,5)$, $(7,7)$, and $(11,11)$ shown in Fig.~\ref{fig:Re}(b), 
are present with relative intensities of $\pm 0.92$. 
All other components have relative intensities smaller than $0.3$. 
Note that inversion symmetry is preserved at the midpoint of the $(4,7)$ and $(5,11)$ bonds.
In deriving the numerical results in Fig.~\ref{fig:Re},
we put $T=0.01~\rm{eV}$ and the other model parameters as $n=11.2$, $\alpha_s =0.9$, $\phi =0.08~\rm{eV}$.
Note that the quantitative numerical results depends on the model parameters.


\end{document}